\documentclass[usenatbib,useAMS,usegraphicx]{mn2e}

\usepackage{subfig}

\newcommand{\agl}{AGL~J2241+4454}
\newcommand{\hd}{HD~215227}
\newcommand{\qsof}{3FGL~J2201.7+5047}
\newcommand{\msun}{$M_{\sun}$}

\newcommand{\pcm}{ph~cm$^{-2}$~s$^{-1}$}
\newcommand{\fermi}{\textit{Fermi}}
\newcommand{\agile}{\textit{AGILE}}
\newcommand\aap{A\&A}

\newcommand\aapr{A\&ARv}

\newcommand\apj{ApJ}
\newcommand\apjs{ApJS}
\newcommand\apjl{ApJ}
\newcommand\araa{ARA\&A}
\newcommand\mnras{MNRAS}
\newcommand\nat{Nature}

\title[An updated gamma-ray analysis of the Be-BH binary HD~215227]{An updated gamma-ray analysis of the Be-BH binary HD~215227}

\author[Michael J. Alexander and M. Virginia McSwain]{Michael J. Alexander$^{1}$\thanks{E-mail: mia313@lehigh.edu} and M. Virginia McSwain$^{1}$\\
$^{1}$Lehigh University, 16 Memorial Drive East, Bethlehem, PA 18018, USA}

\begin{document}

\date{Accepted 2015 February 21.  Received 2015 January 29; in original form 2014 August 5}

\pagerange{\pageref{firstpage}--\pageref{lastpage}} \pubyear{2014}

\maketitle

\label{firstpage}

\begin{abstract}
We report an updated analysis of the gamma-ray source \agl\ that was detected as a brief two-day flare in 2010 by the \agile\ satellite. The high-energy emission of \agl\ has been attributed to the binary system \hd, which consists of a Be star being orbited by a black hole making it the first known Be-black hole binary system. We have analyzed the \agile\ data and find a gamma-ray flux of $(1.8\pm0.7)\times10^{-6}$ \pcm, in agreement with the initial report. Additionally, we examined data from the \fermi\ LAT over several time intervals including the two day flare, the folded orbital phase, and the entire mission ($\sim$6-years). We do not detect \agl\ over any of these time periods with \fermi\ and find upper limits of $1.1\times10^{-7}$ \pcm\ and $5.2\times10^{-10}$ \pcm\ for the flare and the full mission, respectively. We conclude that the \hd\ Be-black hole binary is not a true gamma-ray binary as previous speculated. While analyzing the \fermi\ data of the \agl\ region, we discovered a previously unknown gamma-ray source with average flux of $(13.56\pm0.02)\times10^{-8}$ \pcm\ that is highly variable on monthly timescales. We associate this emission with the known quasar 87GB 215950.2+503417.
\end{abstract}

\begin{keywords}
binaries: general -- gamma-rays: stars -- stars: black holes -- stars: emission-line, Be -- stars: black holes -- quasars: general
\end{keywords}

\section{Introduction}
Microquasars are typically binary systems consisting of a normal (non-degenerate) star from which material is pulled into an accretion disk around a black hole (BH) companion, though in some cases the accreting object may be a neutron star (NS; see \citealt{z13} for a review of microquasars). These objects may be considered as scaled down versions of extra-galactic quasars as they appear to share similar properties. They emit at a wide range of frequencies from radio through gamma-ray, where the emission at different energies is often correlated through disk-jet interactions \citep{rm06}. The accretion disk in microquasars can be fed from Roche-lobe overflow of the companion \citep{mr86} or from dense the stellar winds of a massive secondary star \citep{vcg92}.

The objects Cyg X-1 and Cyg X-3 are known microquasars and high-mass X-ray binaries (HMXB). While Cyg X-1 is known to harbor a 15 $M_{\odot}$ BH \citep{oma11}, the nature of the compact object in Cyg X-3 is remains open, though some evidence suggests a 2 -- 5 $M_{\odot}$ BH companion \citep{sts10,zmb13}. Both systems have been shown to emit gamma-ray flares \citep{bpl10,bps11,btp13}, with some evidence for orbital modulation in Cyg X-3 \citep{aaa09b}. The gamma-ray emission is associated with the soft X-ray state in Cyg X-3 \citep{btc12}, the hard X-ray state in Cyg X-1 \citep{mzc13}, with heightened radio emission in both. 

Another class of objects are the gamma-ray binaries of which five are known: LS I +61\degr303, LS 5039, PSR B1259-63, HESS J0632+057, and 1FGL J1018.6-5856 \citep{d13}. These objects exhibit high-energy spectra that peak at GeV and TeV energies, in contrast to X-ray binaries whose spectra at keV energies. It is suspected that the compact companions of gamma-ray binary systems are NS and that the gamma-ray emission arises from the colliding pulsar and stellar winds \citep{d13}. However, gamma-ray emision may also arise from the interaction between a clumpy stellar wind and the jet produced by a microquasar \citep{ort09}.

In 2010, the \agile\ satellite \citep{tba09} detected a gamma-ray flare with a flux of $1.5\times10^{-6}$ \pcm\ ($>$100 MeV) with a significance greater than $5\sigma$ from a previously unknown source, \agl\ \citep{lvs10}. The location of the flare coincides with a known Be star, \hd\ (=MWC 656; \citealt{mb43}), whose binary nature was elicited only after the discovery of the gamma-ray flare \citep{wgm10}. This began speculation that \agl\ could be the sixth identified gamma-ray binary. \citet{mkn13} performed an analysis of \fermi\ data and found upper limits (90\% C.L.) of $9.4\times10^{-10}$ \pcm\ for the 3.5-year average flux and $7.2\times10^{-8}$ \pcm\ for the flare. They also investigated orbital modulation by folding the data over the period of the binary and found upper limits for each phase bin in the neighborhood of $10^{-8}$ \pcm. So far, observations from \fermi\ have been unable to confirm the detection of \agl.

Recently, \citet{cnr14} found the Be star's companion to be a stellar-mass BH, making it the first Be star in a black hole binary (BHB) system and bringing the total number of confirmed BHBs to 25 \citep{z13}. The secondary Be star has been classified as B1.5--2 III \citep{cnr14}, while a previous analysis found a spectral type of B3 IVne+sh \citep{wgm10}. The measured orbital period of 60.37 days and Be star mass of 10--16 $M_{\odot}$ yield a BH mass of 3.8--6.9 $M_{\odot}$ \citep{cnr14}. \hd\ has also been identified as a source of X-ray emission by \citet{mpr14}, who find a non-thermal X-ray luminosity $L_{X}=3.7\times10^{31}$ erg~s$^{-1}$ corresponding to $3.1\times10^{-8}~L_{\mathrm{Edd}}$ for a 3.8--6.9 \msun\ BH, consistent with the previously found upper limit of $L_X < 1.6\times10^{7} L_{\mathrm{Edd}}$ \citep{cnr14}. This low luminosity is in stark contrast to Cyg X-1, which has an $L_{X}\approx10^{37}$ erg~s$^{-1}$ and $L_{\mathrm{Edd}}\approx10^{-2}$ \citep{gmf12}. This suggests that the system was in a quiescent state at the time of the X-ray observations. While there is a strong contrast at X-ray energies, neither Cyg X-1 nor \agl\ show evidence for sustained gamma-ray emission, but have both been detected as gamma-ray flares by \agile\ \citep{lvs10,bpl10}. Three separate \agile\ detections of Cyg X-1 were confirmed as low-significance counterparts in \fermi\ data \citep{btp13}. The highest significance detections were more likely to be found during transitional stages, perhaps coinciding with jet formation or destruction.

The low X-ray and radio flux from \hd\ places near the low-mass X-ray binary (LMXB) A~0620-003 ($L_{X} = \sim10^{30.5}$ erg~s$^{-1}$; \citealt{gmn01}) on the well-formed radio/X-ray luminosity relation \citep{mpr14}. The X-ray luminosity of \hd\ ($\sim10^{-8} L_{Edd}$; \citealt{mpr14}) is also found to be similar to the median value of sample of eight quiescent systems ($L_{X} = 5.5\times10^{-7} L_{Edd}$; \citealt{rrm14}). In this quiescent state, the X-ray luminosity of these may be dominated by emission from an advection-dominated accretion flow (ADAF) rather than jet interactions \citep{rrm14}.

In light of new evidence for \agl\ as both a HMXB and a BHB, we present an updated analysis of gamma-ray emission towards \agl\ including 6-years worth of \fermi\ data. We aim to determine whether or not \agl\ belongs in the rare class of gamma-ray binaries. Section~\ref{agile-sec} presents the details of our analysis of \agl\ using \agile\ data, Section~\ref{fermi-sec} includes updated analysis of $\sim$6 years of \fermi\ data, Section~\ref{qso-sec} highlights the detection of a newly identified gamma-ray point source in the nearby field, and Section~\ref{disc-sec} discusses and summarizes our results.

\section{\agile\ Analysis of \agl}
\label{agile-sec}
The \agile\ satellite mission launched in 2007 by the Italian Space Agency to survey the sky from hard X-rays through gamma-rays \citep{tba09}. The main instrument for detecting gamma-rays is the Gamma-Ray Detecting Imager (GRID), which is sensitive in the energy range from 30 MeV -- 50 GeV. The GRID PSF containment radius is 1--2\degr\ at 300 MeV with an error box radius of 6--20\arcmin\ depending on source intensity and spectral shape. One of the main science drivers is rapid response to gamma-ray bursts and other high energy transient phenomena, so \agile\ was designed for high precision timing and is capable of detecting strong ($>10^{-6}$ \pcm) point sources with high significance in just 1--2 days of observation.

\citet{lvs10} reported the gamma-ray flare \agl\ at the position $(l=100.0$, $b=-12.2)$ $\pm\ 0\fdg6$ (stat.) $\pm\ 0\fdg1$ (syst.) lasting from 25 July 2010 01:00 UTC through 26 July 2010 23:30 UTC. The measured flux during this time was $1.5\times10^{-6}$ \pcm\ at 5$\sigma$. To analyze this source we downloaded data covering that time period from the ASI Data Science Center website\footnote{http://agile.asdc.asi.it/}. Data reduction followed the GRID analysis manual for using the latest version of the \agile\ software (v5.0).

The position, flux, and spectral index of \agl\ were set to the initial discovery values and allowed to vary during the maximum likelihood fitting. The resulting fit yield a position of $(l=100\fdg1, b=12\fdg2)$, a flux of $(1.2\pm0.6)\times10^{-6}$ \pcm, and a spectral index of $-1.3\pm0.4$ with a significance of $3\sigma$. Because of the low overall photon count ($7\pm4$), we also performed the analysis with the spectral index fixed to a value of $-2$ to reduce the number of free parameters. This yields a flux of $(1.8\pm0.7)\times10^{-6}$ \pcm\ and a total photon count of $11\pm5$ also at $3\sigma$, in agreement with the initially reported value \citep{lvs10}. 

\begin{figure}
\centering
\includegraphics[width=80mm]{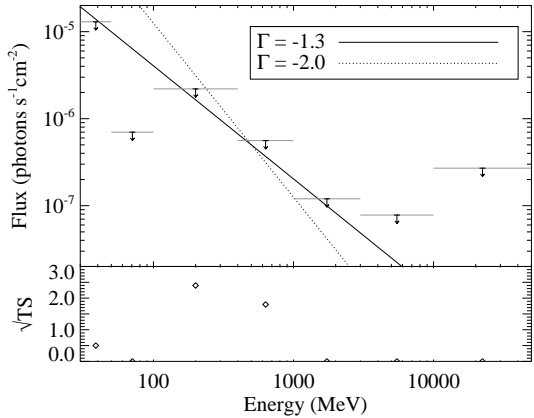}
\caption{Gamma-ray flux upper limits measured by \agile\ for different energy ranges. The horizontal grey lines indicate the bandwidth of each energy range. The solid black line shows a representative power-law with a slope of $\Gamma=-1.3$, while the dotted line is a power-law of slope $\Gamma=-2.0$.}
\label{aglflux}
\end{figure}

We also investigated any spectral dependence by performing the same likelihood analysis on the different energy ranges available from \agile. Figure~\ref{aglflux} shows the measured fluxes broken into several energy bands: 50--100 MeV, 100--400 MeV, 400 MeV--1 GeV, 1--3 GeV, 3--10 GeV, and 10--50 GeV. No indivual band is detected above 3$\sigma$ and all points are represented as upper limits. The solid black line shows a power-law slope of $\Gamma=-1.3$ while the dotted line has a slope of $\Gamma=-2.0$. These lines are representative and not fits as there are no solid detections in the individual energy bands.

\section{Updated \fermi\ Analysis of \agl}
\label{fermi-sec}
The \fermi\ gamma-ray satellite was launched in 2008 by NASA to observe the high-energy sky. On-board are several instruments including the Large Area Telescope (LAT; \citealt{atwood09}). The LAT has a 2.4 sr field of view and covers an energy range from 30 Mev to 300 GeV. We downloaded\footnote{http://fermi.gsfc.nasa.gov/ssc/} and analyzed the Pass 7 Reprocessed data including photons within a 20\degr\ radius around \agl\ and energies from 100 MeV to 300 GeV. All analyses employed the \fermi\ \textsc{science tools} v9r32p5 package as well as the latest Galactic diffuse model (gll\_iem\_v05), isotropic spectral template, instrument response function, and point source catalog to which we added a source at the location of \agl. Following previous studies of other gamma-ray binaries \citep{aaa09a,htt12}, we modeled the source spectrum as a power law plus an exponential cutoff of the form,
\begin{equation}
\frac{dN}{dE} = N_{0}\left(\frac{E}{E_{0}}\right)^{\Gamma}\mathrm{exp}\left[-\left(\frac{E}{E_{\mathrm{cutoff}}}\right)\right],
\end{equation}
where $N_{0}$ is the normalization, $E_{0}$ is the scale factor, $\Gamma$ is the power-law index, and $E_{\mathrm{cutoff}}$ is the high-energy cutoff.

To analyze the flare, we restricted the \fermi\ good time intervals (GTIs) to the 46.5 hr duration of the flare \citep{lvs10}. Of this time, \fermi\ pointing and quality cuts limit the useful observations to 16.8 hrs of data. Our analysis shows no detection of the flare by \fermi, and we derive a 68\% (1$\sigma$) confidence upper limit for the flare of $1.1\times10^{-7}$ \pcm, consistent with \citet{mkn13}. Given the strength of the flare detected by \agile\ ($\sim10^{-6}$ \pcm) it is somewhat surprising that it was not detected, however, the low fraction of \fermi\ observations of the \agl\ flare ($\sim36\%$) may not have been enough time for the source to rise above the background flux. 

We performed a similar analysis for the $\sim$6 years of \fermi\ LAT observations to look long-term gamma-ray emission and find no evidence for sustained emission. We set a 1$\sigma$ upper limit to the average flux of $5.2\times10^{-10}$ \pcm, nearly a factor of two lower than the previous upper limit, which likely owes to the additional data available.

\begin{figure}
\centering
\subfloat{\includegraphics[width=41mm]{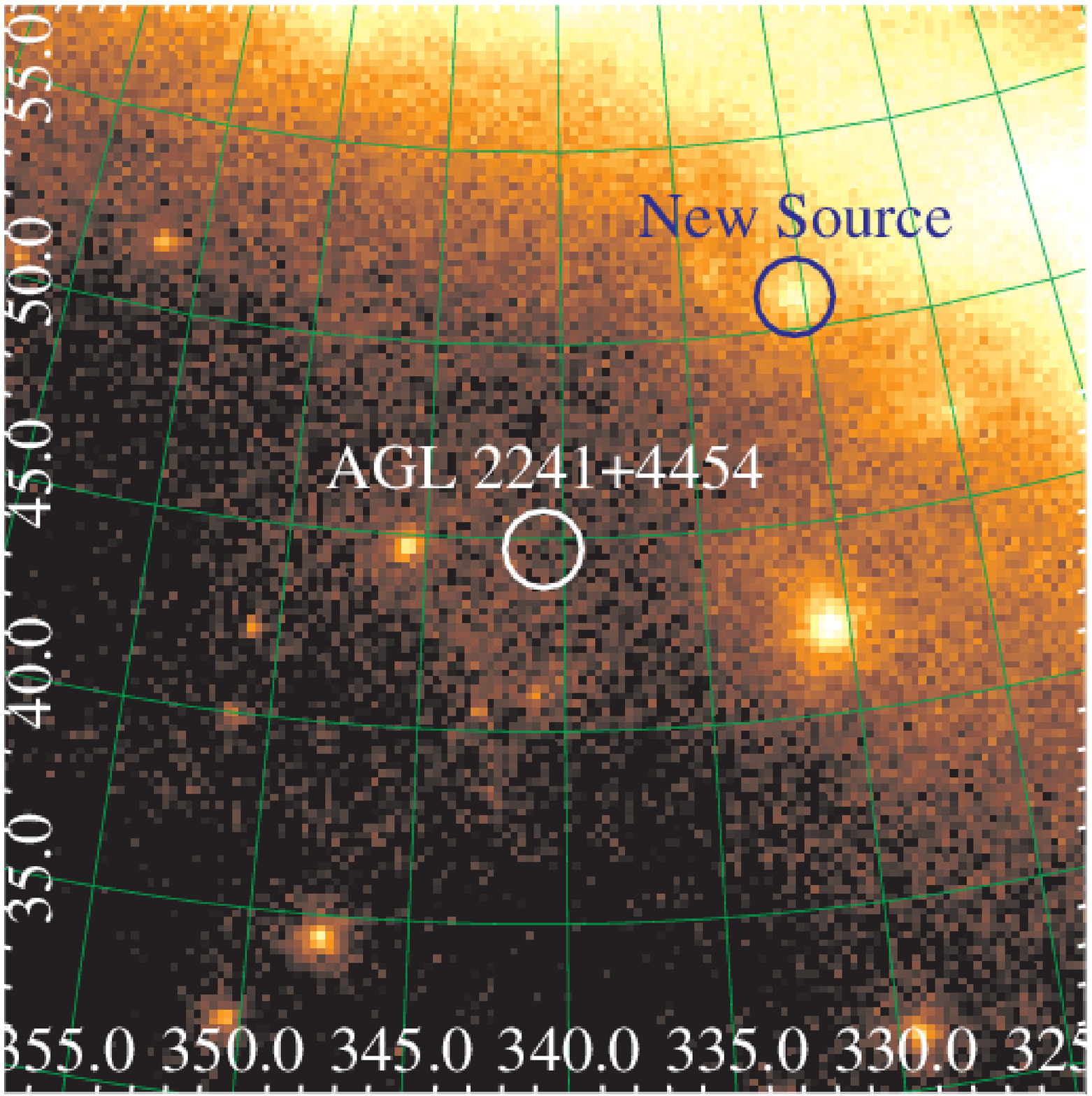}}
%\subfloat{\includegraphics[width=40mm]{hdmodelmap.eps}}
\subfloat{\includegraphics[width=41mm]{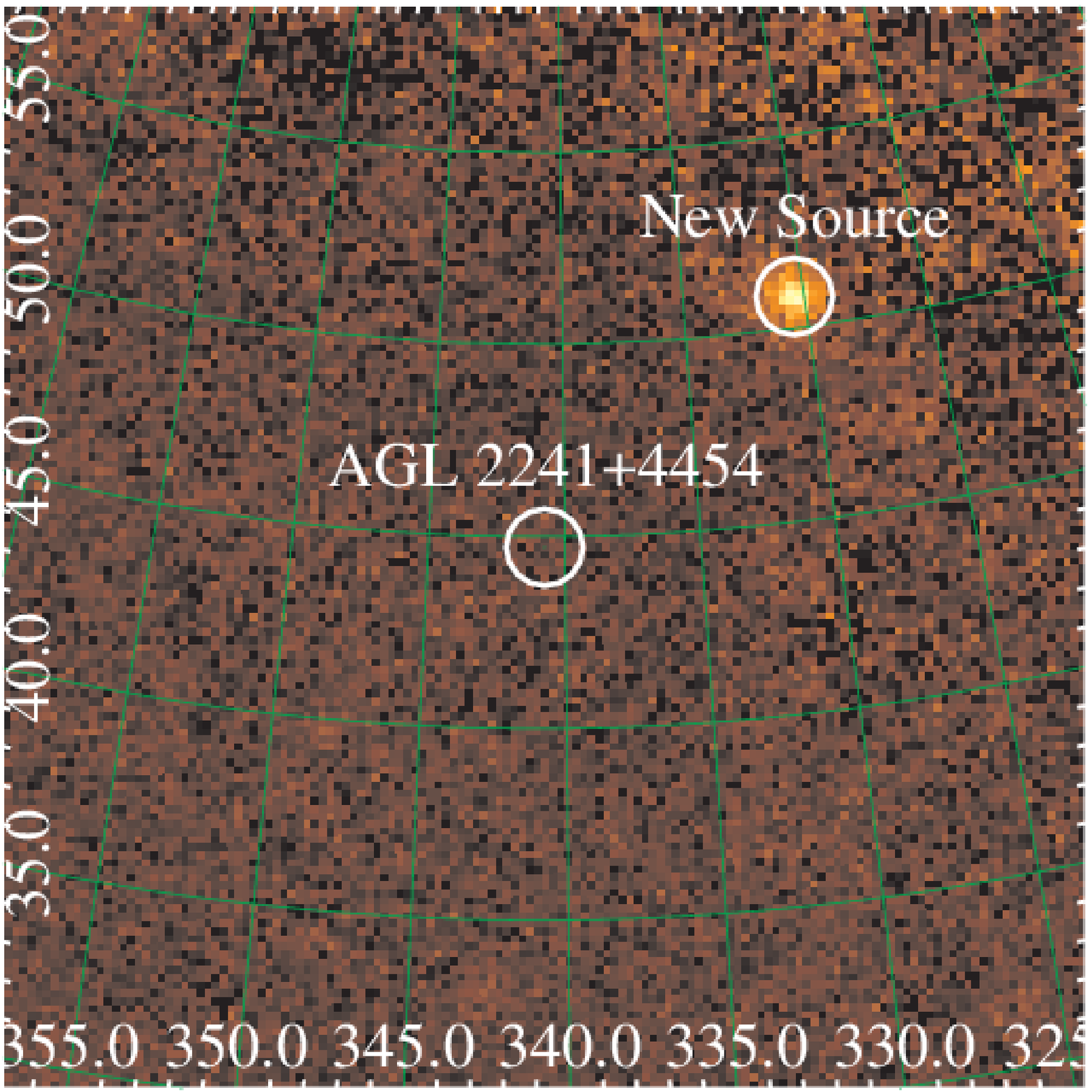}}
\caption{Photon count map for the 6-year dataset (left panel) and the residual map (right panel). The 1\degr\ cyan circles show the location of \agl\ while the dark blue circles mark the previously unknown source \qsof.}
\label{maps}
\end{figure}

Figure~\ref{maps} shows the photon counts map on the left and the residual (counts minus model) map on the right. There is no obvious source of gamma-ray emission at the location of \agl\ above the level of the bright diffuse emission from the Galactic plane. There is an obvious residual leftover in the upper right of the image that is a previously unknown gamma-ray source (Section~\ref{qso-sec}).

\begin{figure}
\centering
\includegraphics[width=7cm]{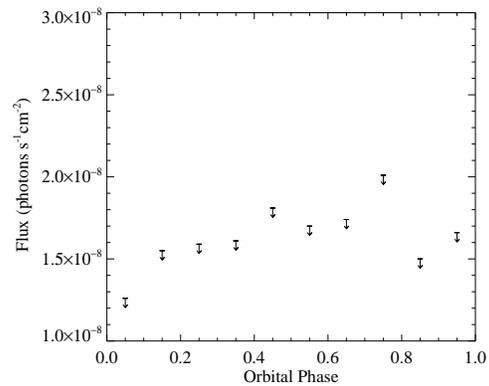}
\caption{Phase-binned gamma-ray fluxes for \agl; all points are 1$\sigma$ upper limits. The estimated \agile\ flux for \agl\ ($1.5\times10^{-6}$ \pcm) is two orders of magnitude larger than the 6.5 yr average phase binned limits.}
\label{hdflux}
\end{figure}

Known gamma-ray binaries such as LS I +61\degr303 \citep{aaa09a} show strong gamma-ray flux modulation over the orbit, so we searched for evidence of this from \agl. We divided the data into time bins of 6.037 days, one-tenth of the orbital period, and folded it over the full time range covered by the data using a reference time of HJD = 2,453,243.7 as phase $\phi=0.0$ \citep{cnr14}. Figure~\ref{hdflux} shows the results of the phase-binned analysis. There are no signficant detections at any orbital phase, and all the points shown are upper limits. 

It is curious that \agl\ was detected by \agile\ and not by \fermi\ since the flux was measured above $10^{-6}$ \pcm. As mentioned earlier, one possible explanation is that 16.8 hrs of on-source time for the LAT was insufficient despite the flare's luminosity. To test this possibility, we ran a likelihood analysis on the source LS~I~+61~303, which has an average flux of $0.9\times10^{-6}$ \pcm\ \citep{htt12}, comparable to that of \agl. LS~I~+61~303 was only observed for a total of 4 hrs during the flare, so we chose a different time that matched the 16.8 hrs of \agl\ observation. For this time period, we find a flux of $0.3\times10^{-6}$ \pcm\ at the 4$\sigma$ level. This situation is not entirely analogous to that of \agl\ as LS~I~+61~303 emits consistently at $\sim10^{-6}$ \pcm, while \agl\ may only have peaked at that flux for fraction of the flare.. However, it demonstrates that the LAT is able to detect bright sources in very short periods time.

\begin{figure}
\centering
\includegraphics[width=7cm]{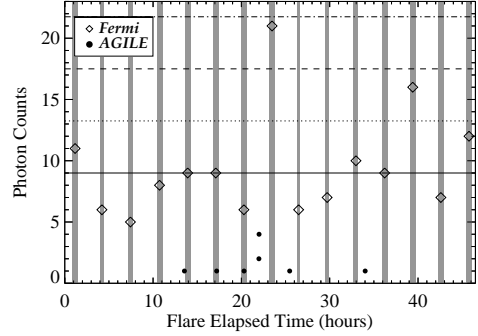}
\caption{Photon counts for \fermi\ (open diamonds) and \agile\ (filled circles) during the flare. The \fermi\ data points represent non-background subtracted photon counts for the GTI (filled grey stripes), while the \agile\ data was binned every 60 seconds. The horizontal lines mark the \fermi\ median (solid), and $1\sigma$ (dotted), $2\sigma$ (dashed), and $3\sigma$ (dot-dashed) above the median. $T_{0}$ is the reported flare start time of 01:00 UTC on 25 July 2010.}
\label{timing}
\end{figure}

Another possible explanation is that the flare was not coincident with \fermi\ GTIs and the flux peak was not actually observed. Figure~\ref{timing} plots the \agile\ and \fermi\ photon arrival times versus the \fermi\ GTIs during the flare; the time begins at the reported flare start. \fermi\ photon counts (open diamonds) are summed for each individual \fermi\ GTI, which are marked by the grey shaded areas. The arithmetic median of the \fermi\ count rate is shown as the solid horizontal line, and simple $1\sigma$, $2\sigma$, and $3\sigma$ standard deviations from the median are the dotted, dashed, and dot-dashed lines, respectively. \agile\ photons are binned every 60 seconds for better time resolution. The data represent all `clean' \agile\ and \fermi\ source plus background photons within 5\degr\ of \hd. For the \agile\ data, we followed the \agile\ users' guide to select the following additional criteria: event status \textit{G}, Earth tolerance 5\degr, and orbital phase code 2. The \fermi\ data were extracted only after passing the photon files through the \textsc{gtmktime} step used for the primary data analysis.

Figure~\ref{timing} shows that the gamma-rays detected by \agile\ arrived at the detector in the middle of the reported time limits with the peak flux near hour 22 (about 23:00 on 25 July). However, of the 11 photons selected by the constraints given above, just two landed within \fermi\ GTIs. The GTI immediately following the \agile\ spike also shows a peak in the \fermi\ photon count, but did not reach the nominal $3\sigma$ limit. The figure cannot definitively demonstrate the presence/absence of a flare, however, it illustrates the possibility that the \agl\ flare was not detected by \fermi\ because the telescope was simply not looking. If a flare did occur, then it must have lasted less than about six hours for the resulting flux not to have been seen in at least one additional GTI, including the maximum at 23.5 hours. A six-hour window translates to a light-travel distance of 43 AU, which would be an upper limit on the size of the emitting region.

\section{The new gamma-ray source \qsof}
\label{qso-sec}

\begin{figure}
\centering
\includegraphics[width=7cm]{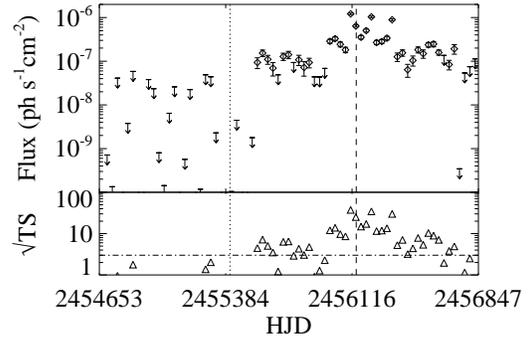}
\caption{Fluxes (top) and significance values (bottom) for \qsof\ with each time bin representing one month of \fermi\ observations. Diamonds represent valid ($>3\sigma$) detections while arrows indicate 1$\sigma$ upper limits. The vertical dotted and dashed lines indicate the time of the last data included for sources in the 2FGL and 3FGL source catalog, and the horizontal dot-dashed line in the bottom panel marks the $3\sigma$ detection limit.}
\label{qsoflux}
\end{figure}

During our initial analysis of \agl, we found a bright object in the residual image (Figure~\ref{maps}, right panel). This source, which had no associated 2FGL source but appears in the new catalog as \qsof, we modeled as a log-parabola to the model file of the form,
\begin{equation}
\frac{dN}{dE} = N_{0}\left(\frac{E}{E_{b}}\right)^{-(\alpha+\beta log(E/E_{b}))} ,
\end{equation}
where $N_{0}$ is the normalization, $E_{b}$ is the break energy, and $\alpha$ and $\beta$ are power-law indices. We then followed the same binned likelihood analysis as before. The source was detected over the energy range 100 MeV -- 300 GeV at $47\sigma$ with a flux of $(13.56\pm0.02)\times10^{-8}$ \pcm, $\alpha=2.577\pm0.004$, $\beta=0.098\pm0.002$, and $E_{b}=505.2\pm0.7$ GeV, while the \fermi\ catalog reports a $27\sigma$ detection for \qsof\ of $(0.219\pm0.015)\times10^{-8}$ \pcm, roughly two orders of magnitude lower. The main cause of the difference is most likely that the source remained in a high flux state ($\ga10^{-7}$ \pcm) for an 8--12 months after the data cutoff date for the 3FGL catalog (Figure~\ref{qsoflux}), which lead to a lower average flux. Also, the \qsof\ was modeled as a power law using a smaller energy range (1--100 GeV) than for our analysis. These are the main factors that make the fluxes different, but not necessarily in conflict. Because many known gamma-ray sources are variable, we then divided the data into 30 day time bins to measure the changes in the high-energy flux.

Figure~\ref{qsoflux} shows the source flux since the start of \fermi\ science operations where each data point represents one month of observations. The dotted and dashed horizontal lines represent the time of last data included in the 2FGL (01 August 2010 at 01:17 UTC) and 3FGL (31 July 2012 at 22:46 UTC) catalogs, respectively. During the time range included in the 2FGL catalog (dashed horizontal line), the detection significance never rose above $3\sigma$ in a given month, which is likely the reason the source was not identified. It appears to have brightened suddenly near the start of 2011 and remained strong through the end of 2013. However,the last few months of data suggest that the object may be entering a low flux state. Continued monitoring is necessary to see if the gamma-ray flux continues to decline.

We used the procedure \textsc{gtfindsrc} to localize the position of the source to $\alpha=330\fdg412$, $\delta=50\fdg828$ with an error circle radius of $0\fdg016$ (57\arcsec). The values agree well with the 3FGL coordinates, $\alpha=330\fdg434\pm0\fdg026$, $\delta=50\fdg800\pm0\fdg025$, where the errors are at the 68\% confidence level. We report errors that are about two times smaller than the \fermi\ catalog, which may also be related to the additional high flux data in our analysis leading to a stronger detection and better localization. We searched the SIMBAD Astronomical Database\footnote{http://simbad.u-strasbg.fr/simbad/} around this position for possible progenitors and found a known blazar, 87GB 215950.2+503417 \citep{gc91}, 62\arcsec\ away. This association was also recently identified by the \fermi\ team in the 3FGL catalog. The quasar is also coincident with X-ray source 1RXS 220140.8+504941 from the ROSAT All Sky catalog \citep{vab99}), which has a count rate of $(2.84\pm0.83)\times10^{-2}$ s$^{-1}$. Prior to its launch, 87GB 215950.2+5031417 was also identified a the catalog of blazars likely to be detected by \fermi\ (\#657; \citealt{srm05}). 

\section{Discussion}
\label{disc-sec}

We have analyzed the \agile\ gamma-ray point source, \agl, to search for evidence that the emission originates from the Be-BH binary system \hd. We find a gamma-ray flux of $(1.8\pm0.7)\times10^{-6}$ \pcm\ at $3\sigma$ significance, in agreement with the initial report. Despite the large \agile\ flux, we find no evidence of emission in the \fermi\ data using three different time binning variations.

There are two main reasons that may be partially or entirely responsible for the \fermi\ non-detection. The first is that the flare was produced mostly lower energy photons ($<$ a few hundred MeV) with a steep power law. The photons detected by \agile\ were mostly in the 100 MeV range where the performance of the LAT suffers from a lower effective area and larger PSF compared to energies higher than 1 GeV. Second, the flare may have been short (less than $\sim$6 hours) and peaked in between \fermi\ GTIs. This theory is supported by Figure~\ref{timing}, which shows that the peak detected by \agile\ occurred outside of the \fermi\ GTIs for the region around the flare. Considering these possibilities, it seems that some combination of the first two is most likely. This issue may be clarified when the \fermi\ science team releases the new Pass 8 data in mid-2015. The improved event reconstruction will provide significant increases in the effective area and field of view, especially at energies below 100 MeV. An updated analysis with the Pass 8 data may at least rule out the possibility that the photons were of too low energy for \fermi\ to detect.

In any case, it appears that the proposed gamma-ray source \agl\ does not fit within the category of known gamma-ray binaries such as LS 5039 or LS~I~+61~303. The five known gamma-ray binaries all show orbital phase modulation of their high-energy (GeV) and very-high energy (TeV) gamma-ray emission \citep{d13}, while \agl\ does not emit gamma-rays at a continuously detectable level, if at all. Instead, it is more likely that the \hd\ binary system is a HMXB and microquasar that has not been discovered earlier owing to its seemingly persistent quiescent state. Deeper X-ray observations will be of future importance to determine whether or not an ADAF model accurately describes \hd\ or if other emission mechanisms are at play. Additionally, contemporaneous X-ray and radio observations may be able to further test disk-jet coupling theories for the system.

During the investigation into \agl, we identified a gamma-ray point source that was not present in the 2FGL catalog, but now appears in the latest 3FGL catalog.  The source, \qsof, is variable on monthly (and perhaps shorter) timescales and has been bright ($>10^{-8}$ \pcm) for the past few years. We, and independently, the \fermi\ LAT team, identify the flat-spectrum radio quasar 87GB~215950.2+503417 as the most probable source of the gamma-ray emission because of its proximity to the gamma-ray peak, its known X-ray and radio emission, and its prior identification as a potential gamma-ray emitter.

\section*{Acknowledgments} 
The Fermi LAT Collaboration acknowledges generous ongoing support from a number of agencies and institutes that have supported both the development and the operation of the LAT as well as scientific data analysis. These include the National Aeronautics and Space Administration and the Department of Energy in the United States, the Commissariat \`{a} l’Energie Atomique and the Centre National de la Recherche Scientifique/Institut National de Physique Nucl\`{e}aire et de Physique des Particules in France, the Agenzia Spaziale Italiana and the Istituto Nazionale di Fisica Nucleare in Italy, the Ministry of Education, Culture, Sports, Science and Technology (MEXT),High Energy Accelerator Research Organization (KEK) and Japan Aerospace Exploration Agency (JAXA) in Japan, and the K. A. Wallenberg Foundation, the Swedish Research Council and the Swedish National Space Board in Sweden. We are grateful for support from the National Science Foundation through grant AST-1109247. This research has made use of the SIMBAD database, operated at CDS, Strasbourg, France. Part of this work is based on archival data, software, or online services provided by the ASI SCIENCE DATA CENTER (ASDC). We also thank the anonymous referees for many helpful comments.

%\bibliographystyle{mn2e} % citation style
%\bibliography{references}

\label{lastpage}

\end{document}